\documentclass[twocolumn,showpacs,preprintnumbers,amsmath,amssymb]{revtex4}

\usepackage{graphicx}
\usepackage{dcolumn}
\usepackage{bm}
\usepackage{booktabs}

\begin{document}
\preprint{APS/123-QED}

\title{THz-range free-electron laser   ESR spectroscopy:\\ techniques and applications in high magnetic fields}
\author{S. A. Zvyagin}
\email{s.zvyagin@fzd.de}
\homepage{http://www.fzd.de/hld}
\affiliation{Dresden High Magnetic Field Laboratory (HLD),
Forschungszentrum Dresden-Rossendorf (FZD), 01314 Dresden, Germany}
\author{M. Ozerov}
\affiliation{Dresden High Magnetic Field Laboratory (HLD),
Forschungszentrum Dresden-Rossendorf (FZD), 01314 Dresden, Germany}
\author{E. \v{C}i\v{z}m\'{a}r\footnote{Present address: Centre of Low Temperature Physics of the Faculty of Science of P.J. \v{S}\'{a}farik University and IEP SAS, Park Angelinum 9, SK-04154 Ko\v{s}ice, Slovakia}}
\affiliation{Dresden High Magnetic Field Laboratory (HLD),
Forschungszentrum Dresden-Rossendorf (FZD), 01314 Dresden, Germany}
\author{D. Kamenskyi}
\affiliation{Dresden High Magnetic Field Laboratory (HLD),
Forschungszentrum Dresden-Rossendorf (FZD), 01314 Dresden, Germany}
\author{S. Zherlitsyn}
\affiliation{Dresden High Magnetic Field Laboratory (HLD),
Forschungszentrum Dresden-Rossendorf (FZD), 01314 Dresden, Germany}
\author{T. Herrmannsd\"{o}rfer}
\affiliation{Dresden High Magnetic Field Laboratory (HLD),
Forschungszentrum Dresden-Rossendorf (FZD), 01314 Dresden, Germany}
\author{J. Wosnitza}
\affiliation{Dresden High Magnetic Field Laboratory (HLD),
Forschungszentrum Dresden-Rossendorf (FZD), 01314 Dresden, Germany}
\author{R. W\"{u}nsch}
\affiliation{Institute of Radiation Physics,
Forschungszentrum Dresden-Rossendorf (FZD), 01314 Dresden, Germany}
\author{W. Seidel}
\affiliation{Institute of Radiation Physics,
Forschungszentrum Dresden-Rossendorf (FZD), 01314 Dresden, Germany}

\date{\today}

\begin{abstract}
The successful use of   picosecond-pulse free-electron-laser (FEL) radiation  for the  continuous-wave  THz-range electron spin resonance (ESR) spectroscopy has been demonstrated.  The  combination of two linac-based  FELs (covering  the wavelength range of 4 - 250 $\mu$m) with pulsed magnetic fields up to 70 T allows for  multi-frequency ESR spectroscopy in a frequency range of 1.2 - 75 THz  with a spectral resolution better than 1$\%$.
The performance of the spectrometer is illustrated with  ESR spectra obtained in the 2,2-diphenyl-1-picrylhydrazyl (DPPH)  and the low-dimensional organic material (C$_6$H$_9$N$_2$)CuCl$_3$.
\end{abstract}

\maketitle
\section{\label{sec:level1}Introduction}

Electron spin resonance (ESR; sometimes called electron paramagnetic resonance or EPR) provides a powerful means for the investigation of  low-energy magnetic excitations in numerous magnetic substances.  It is widely accepted in the high-field ESR community that despite of  high resolution and sensitivity, the application range of commercially available ESR spectrometers (with the highest frequency of $\sim$0.1 THz and magnetic fields up to 6 T) appears  considerably limited. The use of such one-frequency spectrometers becomes rather inefficient when studying magnetic systems with a large zero-field splitting, complex frequency-field diagrams of magnetic excitations or materials with  field-dependent  ESR parameters ($g$-factor, line-width, etc.). Multi-frequency  ESR techniques   employing several frequencies and combined with high magnetic fields allows for more detailed ESR studies of  such systems. This kind of ESR spectroscopy   has a broad range of applications in  solid-state physics, material science,   and chemistry (see for instance Ref. \cite{Hassan-spectrometer,Hill,Katsumata,Nojiri,Ohta,Zvyagin,Krzystek,Zvyagin2}).

ESR spectroscopy is based on the detection of resonance absorption of electromagnetic radiation corresponding to transitions between  electron-spin energy levels split by internal effects (crystal-field anisotropy, exchange interactions, etc.) and/or by an applied magnetic field. The strength of such interactions  in magnetic materials  can vary from  tens of millikelvin  to hundreds and even thousands of Kelvin (which corresponds to a frequency range from tens of MHz to hundreds of THz). That is why extending the frequency range of  ESR techniques  beyond 1 THz appears to be   one of the central issues in  modern ESR spectroscopy.

Due to the lack of intensive sources of electromagnetic  radiation the frequency range between 300 GHz and 10 THz  is known as ``THz gap".  It is important to mention that  the output power of conventional sources of  monochromatic  submillimeter-wavelength radiation used for high-frequency ESR (such as backward wave oscillators, BWOs  \cite{Eremenko,Zvyagin-spectrometr,Nojiri-BWO}, Gunn-diodes, and  millimeterwave  vector network analyzers   \cite{Hill-spectrometer,Hagiwara-spectrometer,Kataev-spectrometer}) significantly drops approaching the THz frequency range (down to some hundreds and even tens of $\mu$W). It imposes serious restrictions for ESR in this frequency range. Employment of submillimeter wave molecular-gas lasers  (pumped by a tunable CO$_2$ laser) and quantum-cascade lasers (QCL) in combination with high magnetic fields  has been proven a useful technique   for studying the cyclotron resonance in semiconductors. On the other hand, a rather discrete spectrum of  radiation produced by molecular-gas lasers (which is limited for each gas or mixture of gases),  and the relatively narrow frequency range of QCLs significantly reduces the use of these radiation sources, when  frequency tunability is required. That is why the application of  free-electron lasers (FEL) as powerful radiation sources,  tunable over a large frequency range,    appears  to be very promising.

The first attempt to combine FEL and pulsed magnetic fields was made using the FELIX radiation source at Rijnhuizen \cite{Vaughan,vanBockstal,Langerak,Langerak2}.  Although the idea of employing FELs as THz radiation sources in  ESR spectroscopy seems,  at the first glance, very straightforward, it has remained questionable whether a sufficiently high resolution required for most ESR applications (at least better than 1$\%$) can be achieved.  It is worth to mention that in contrast to radiation  produced by conventional   sources   (for instance,  Gunn-diodes) used for continuous-wave (cw)  ESR spectroscopy, the radiation produced by FELs has a pulse structure, with a typical  pulse duration of the order of  microseconds (in case of electrostatic, or van de Graaf,  accelerator) or picoseconds (linear accelerator, hereafter linac).  Due to the bandwidth-limited (or sometimes called Fourier-limited) pulse nature, the FEL radiation  is not ideally monochromatic - particularly in case of  linac-based FELs. Furthermore, the FEL pulse power can reach hundreds of kW/cm$^2$, which might lead to a number of parasitic effects, including for instance overheating of the sample and even its destruction as result of  ionization defects. Effects of  optical bleaching under the influence of high-power FEL radiation were reported for InAs/GaSb \cite{Vaughan} and InAs/Al$_x$Ga$_{1-x}$Sb  \cite{Singh-saturation} structures, opening new opportunities in the so-called saturation spectroscopy \cite{Helm-saturation}  but, on the other hand, demonstrating a potential problem when employing high-power pulsed radiation produced by FELs for the conventional ``low-power'' spectroscopy.

Here,  we report on the  successful application of the linac-based FEL at the Research Center Dresden-Rossendorf (Forschungszentrum Dresden-Rossendorf, FZD) as  a tunable source of THz radiation for high-field ESR.  The spectrometer operates in the quasi-cw mode and allows for  ESR experiments in the wavelength range of 4 - 250 $\mu$m (which corresponds to  1.2 - 75 THz or 40 - 2500 cm$^{-1}$ in frequency, and 5 - 310 meV in energy) in pulsed magnetic fields up to 70 T with a spectral resolution better than 1$\%$.  The remainder of this paper is organized as follows. In section II, an overview of  FEL-based THz-radiation sources at the FZD is presented. Section III describes the pulsed-field user facilities  at the  Dresden High Magnetic Field Laboratory (Hochfeld-Magnetlabor Dresden,  HLD) used for high-field ESR spectroscopy.  In section IV, the description  of the FEL-based spectrometer is given. In conclusion, we discuss further potential  of FEL spectroscopy, including those in high magnetic fields.

\section{\label{sec:level1}Radiation source FELBE}
\
FELBE is the acronym for the free-electron laser (FEL) facility at the superconducting Electron Linear accelerator of high Brilliance and Low Emittance (ELBE) located at the FZD in Dresden. There are two FELs at the FZD. Both of them are  Compton FELs  working in the upper region of the low-gain regime (G $\sim$ 10 - 80 $\%$).
A mid-infrared (IR)  FEL (undulator U27) can be operated  in the wavelength range of 4 - 22 $\mu$m \cite{Michel-FEL}, using an electron beam  energy varying from 15 to 35 MeV.   A long-wavelength far-IR  FEL (undulator U100) is equipped with a partially waveguided resonator  and
can be operated  in the wavelength range of 18 - 250 $\mu$m \cite{Lehnert-FEL}.  Thus,  the combination of two FELs allows to quasi-continuously cover the wavelength range from 4 to 250 $\mu$m.  The pulse energy  ($\leq5\mu$J)  depends  on the radiation wavelength, electron-beam energy, and resonator and undulator parameters.  Driven by a superconducting linac, FELBE continuously generates IR radiation pulses with a repetition rate of 13 MHz. The possibility to operate  in the cw regime is one of the most important advantages of FELBE at the FZD. This regime is of particular importance for pulsed-field ESR (with a typical magnetic field-pulse duration from 10 ms to some hundreds of ms), allowing to avoid complex problems of synchronization of FEL radiation and  magnetic-field pulses, permitting longer acquisition times and thus better signal-to-noise ratio.  In addition, operation in  macropulse  (macrobunch duration $\geq$ 100$\mu$s, repetition rate $\leq$ 25 Hz) as well as  single-pulse modes are possible. High-level linear polarization of the FEL radiation ($\approx98\%$) was confirmed experimentally \cite{polarization}. In Table 1,  parameters of both FELs are summarized.

\begin{table}[hbt]
   \centering
   \caption{FEL specifications}
   \begin{tabular}{lcc}
   \hline
       \toprule
       \textbf{Parameter} & \textbf{U27} & \textbf{U100} \\
       \hline
       \midrule
         Undulator period (mm)              & 27.3             & 100        \\
         Number of periods N$_u$          & 2x34             & 38        \\
         Undulator parameter K$_{rms}$        & 0.3 - 0.7          & 0.5 - 2.7  \\
         Wavelength ($\mu$m)        & 4 - 22             & 18 - 250        \\
         Pulse durations (ps)        & 0.9 - 4             & 5 - 30        \\
         Max. pulse energy ($\mu$J)       & 2             & 5        \\
         Max. aver. power (W)     & 30    & 65  \\
       \bottomrule
       \hline
   \end{tabular}
   \label{l2ea4-t1}
\end{table}

\begin{figure}[h]
\begin{center}
\includegraphics[width=0.45\textwidth]{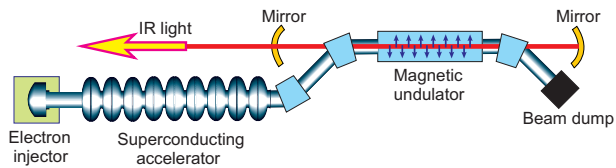}
\caption{\label{fig:felscheme-general} (Color online) Schematic view of one of the FELs used at the FZD.
}
\end{center}
\end{figure}

The FELs at the FZD consist of the following main parts:  a superconducting accelerator (equipped with  electron injector, beam lines, and beam dumps),  magnetic undulator,  and optical resonator (Fig.\ \ref{fig:felscheme-general}).  The electron beam (with up to 1 mA beam current at 35 MeV) is produced by  a radiofrequency (RF) linear accelerator.  The RF power comes from 10 kW clystrons controlled by a low-level RF system.  Electrons are first  accelerated in a  250 keV-thermionic DC electron gun and prebunched in a two-stage RF buncher section. The main acceleration is achieved in two 20 MeV 1.3 GHz linear accelerator modules equipped with cavities made from superconducting Nb and cooled  to 1.8 K. Apart from driving the FELs, the accelerator can be used for generating various kinds of secondary radiation (x-ray, positrons, neutrons, and Bremsstrahlung) \cite{Gabr}.

\begin{figure}[h]
\begin{center}
\includegraphics[width=0.45\textwidth]{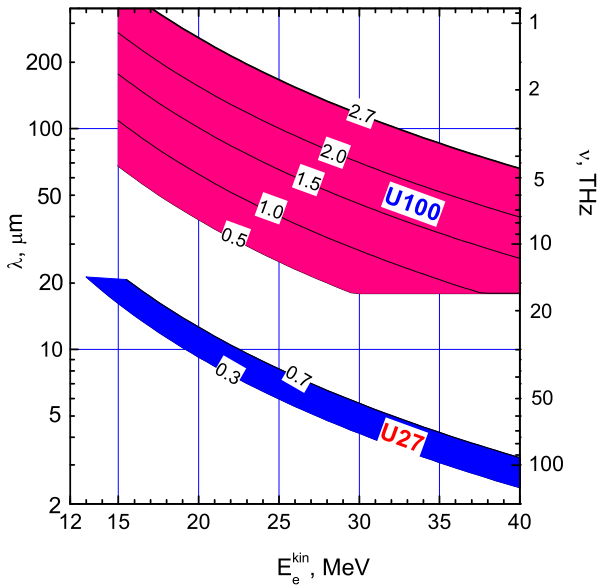}
\caption{\label{fig:energy-range} (Color online) The  wavelength (frequency)  range  of the U27  and U100  undulators   as a function of  kinetic electron energy, $E_e^{kin}$, calculated for different  undulator parameters $K_{rms}$. }
\end{center}
\end{figure}

\begin{figure}
\begin{center}
\includegraphics[width=0.46\textwidth]{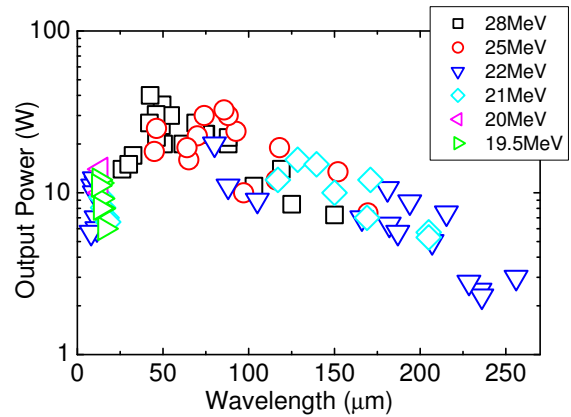}
\caption{\label{fig:Power} (Color online) Typical measured output power vs wavelength for the two undulators, U27 ($\lambda< 20~\mu$m) and U100 ($\lambda>20~\mu$m), for different kinetic electron energies (user statistics for years 2008 - 2009).
}
\end{center}
\end{figure}

The wavelength of the FEL radiation, $\lambda_n$, is determined by the electron energy $\gamma$ (in units of electron rest mass),  the undulator parameter $K_{rms}$  (rms = root mean square), and  the undulator period $\lambda_U$ ($n$ is the radiation harmonic number):
\begin{equation}
\label{lambda}
\lambda_n = \frac{\lambda_U
(1+K_{rms}^2)}{2n \gamma^2}.
\end{equation}
The dimensionless undulator parameter $K_{rms}$ is given by
\begin{equation}
\label{K}
K_{rms}=\frac{e B_U \lambda_U}{2 \pi m_e c},
\end{equation}
where $e$ and $m_e$ are the electron charge and mass, respectively, $c$ is the speed of light, and  $B_U$ is the  rms amplitude of the magnetic field on the undulator axis. The undulator parameter  can be varied by changing the undulator gap which determines  the magnetic field along the undulator axis. The wavelength ranges of the U100 and U27 undulators  as a function of a kinetic electron energy, $E_e^{kin}$, calculated for different  $K_{rms}$  are shown in Fig.\ \ref{fig:energy-range}. The measured output power vs wavelength for the undulators  U27  and U100 at different kinetic electron energies (user statistics for years 2008 - 2009 is presented) is shown in Fig.\ \ref{fig:Power}.

\begin{widetext}

\begin{figure}[tb]
\begin{center}
\includegraphics[width=0.82\textwidth]{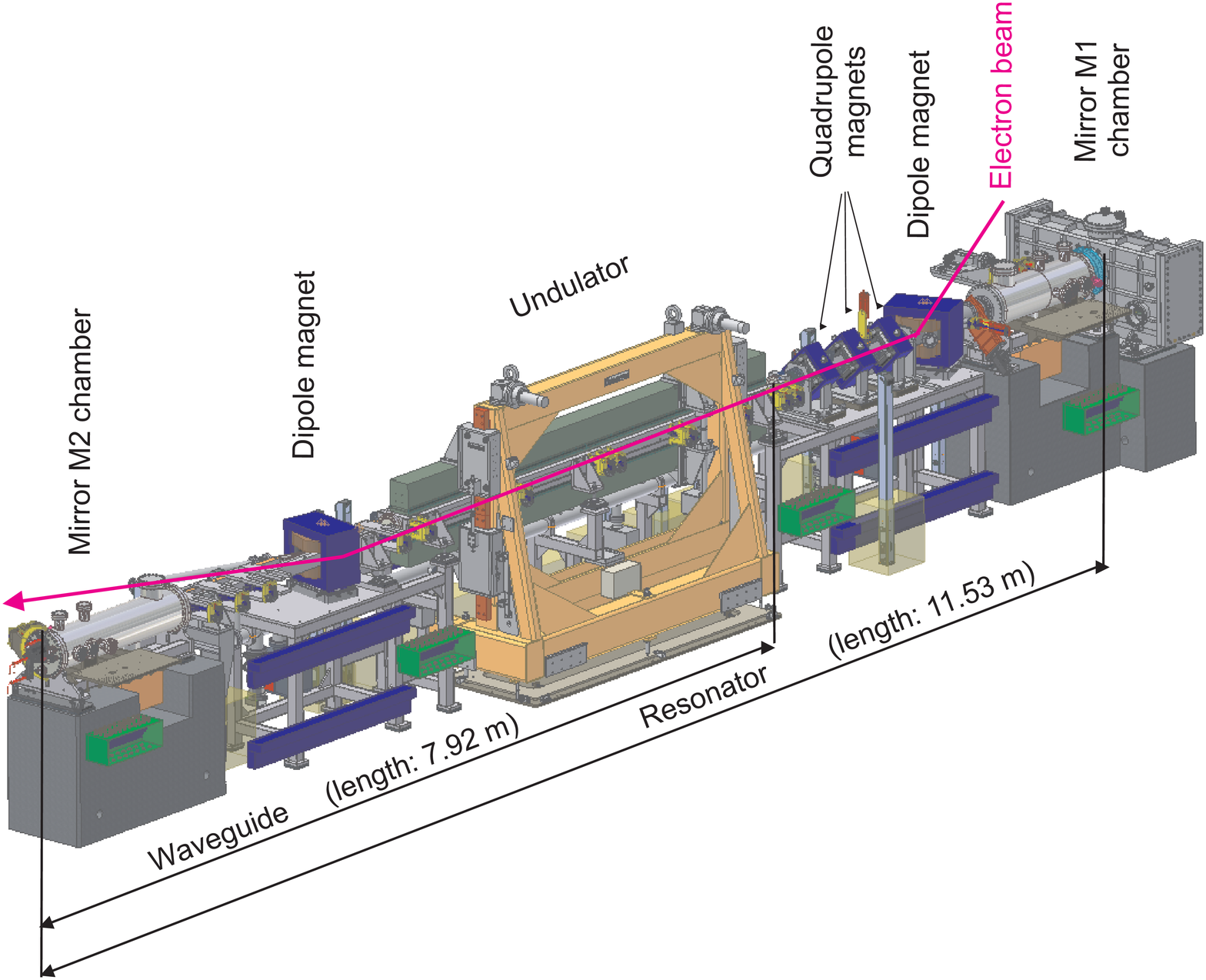}
\caption{\label{fig:FEL2} (Color online) Schematic view of the U100-based FEL. The electron beam enters the FEL from the right side.}
\end{center}
\end{figure}

\end{widetext}

The undulator is the ``heart" of an FEL.  A schematic view of the U100-based FEL (without accelerator section) is shown in Fig.\ \ref{fig:FEL2}. The undulator U100  contains 38 sections of Sm/Co-magnets  and magnet poles (made from soft high-permeability decarbonized iron), with a period of 100 mm (this parameter defines the name of  the undulators, e.g., U100 or U27). The  magnetic field alternates along  the undulator axis with a period  $\lambda_U$ wiggling the electron beam in a direction perpendicular to the undulator axis and the magnetic filed vector.  Due to the transverse modulation  electrons emit light. Another important component of FELs is the  optical  resonator, which for U100 consists of a bifocal upstream (M1) and a cylindrical downstream (M2) mirror. The distance between the two mirrors, $L_R$=11.53 m,  is determined by the repetition rate  of the electron pulses (13 MHz) coming from the accelerator, and chosen to secure an  effective coupling between the electron beam and the oscillating wave in the resonator.
The undulator U100 is equipped with a  partial parallel-plate waveguide (7.92 m long, 10 mm  distance between the plates), confining the IR beam  in  vertical (transverse) direction, increasing the laser gain and allowing a smaller undulator gap (i.e., a higher modulation field).  Because of the large variation of the beam radius determined by the radiation wavelength and expected laser gain,  outcoupling holes of different diameters (2, 4.5 and 7 mm for U100) are  used. The appropriate mirror can be shifted into the right position by using a high-precision linear translation stage.

The undulator U27 has been designed for a  shorter wavelength range (4 to 22 $\mu$m). It is equipped with two planar undulator units with a length of 0.98 m each.  Each unit contains 34 sections of   NdFeB magnets with alternating poles  with a period of 27.3 mm. In contrast to U100, the undulator U27 is not equipped with   a waveguide;  instead just an open optical resonator ($L_R$ = 11.53 m) consisting of two spherical mirrors is used. To optimize the extraction ratio over the whole wavelength range, mirrors with different outcoupling hole sizes (1.5, 2.0, 3.0, and 4.0 mm) are used.  The mirrors are mounted on  a wheel which is fixed on a high-precision rotation stage.

The resonator length can be adjusted and stabilized using a Hewlett-Packard interferometer system. A separate beamline guides the IR beam to a diagnostic station. The radiation wavelength is measured using  a Czerny-Turner-type spectrometer, which contains a turret with three different gratings to cover the wavelength range from 4  to about 230 $\mu$m. A standard Bruker Equinox 55 Fourier-transform spectrometer is used for measuring the spectral radiation parameters at wavelengths up to 250 $\mu$m.

From the diagnostic station,  the beam is guided to user laboratories.  For this purpose, gold-coated copper and stainless-steel mirrors are used. The reflectivity of the mirrors (with a surface roughness being better than  0.2 $\mu$m) is 98.8$\%$ in the entire wavelength range. The IR beam is fed through a 27 m long tunnel  to the adjacent HLD building.   The FELBE-HLD beamline contains 13 mirrors (with effective diameters ranging from 100 to 200 mm). To reduce the number of mirrors to a minimum, some mirrors are used for  beam deflection  as well as for refocusing. To avoid the absorption of IR light by air, the beam has to be guided in pipes which are either evacuated or constantly purged with dry nitrogen gas. The power attenuation by the beamline is   3 - 10 dB and depends strongly on the radiation wavelength and  used  windows. If additional attenuation is required,   the beam power can be  reduced  by 3 - 38 dB using an optical attenuator located at the diagnostic station.  High-level linear polarization of the FEL radiation (better than 95$\%$) after the propagation through the beamline was confirmed experimentally. Further details of the beamline are described in Ref. \cite{Seidel-FEL}.

\begin{figure}[h]
\begin{center}
\includegraphics[width=0.5\textwidth]{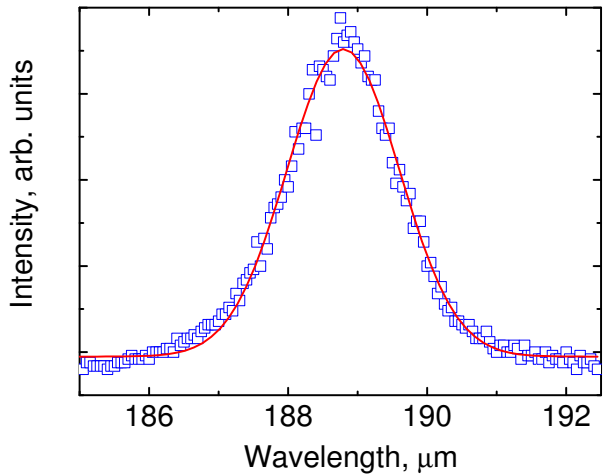}
\caption{\label{fig:FEL-spectrum} (Color online) Example of  spectral  distribution of FEL power (symbols) and the corresponding Gaussian line-shape fit (line) for a wavelength of 188.8 $\mu$m. The full spectral width  at half maximum, $\Delta \lambda / \lambda$, is 0.8$\%$.
}
\end{center}
\end{figure}

The pulse length of the FEL radiation and the spectral distribution of the power was measured by means of an autocorrelator and a spectrometer, respectively \cite{Seidel}. The resulting time-bandwidth product indicates the Fourier-transform-limited operation of both FELs.   The radiation pulse length can be changed by a factor of about 5 by detuning the resonator with respect to its nominal length determined by the electron-bunch repetition rate. As confirmed experimentally,  the spectral width of the produced radiation varies by the same factor. An example of  the spectral  distribution of the FEL power and the corresponding Gaussian line-shape fit  for a wavelength of 188.8 $\mu$m is shown in Fig.\ \ref{fig:FEL-spectrum}. In this example, the full spectral width  at half maximum (FWHM), $\Delta \lambda / \lambda$, of the FEL radiation is 0.8$\%$. This parameter  can be changed continuously, depending on experimental requirements,  approximately from  2 to 0.4$\%$.

\section{\label{sec:level1}Pulsed-field magnets}

The pulsed-field laboratory is equipped with a 50 MJ/24 kV capacitor bank \cite{Herm}.   The power supply consists of  twenty  modules: fifteen modules with a maximum electric energy of 2.88 MJ,  four 1.44 MJ modules, and one 0.9 MJ module.  The modules can be operated in a variety of combinations, depending on technical requirements (including feeding of  multi-coil magnets).

It is important to mention that  pulsed-field magnets are subjects of extremely high electrical, thermal, and mechanical stresses.  The voltage between adjacent layers of the coil can be as high as several kV, requiring special measures for wire insulation.  The coils are cooled by liquid nitrogen. During pulses the magnets undergo a mechanical and thermal shock. Because of the Joule heating, the wire temperature increases from 77 K  to above room temperature within fractions of a second. But the most serious problem is the huge mechanical stress, which arises in the coil at peak field due to the Lorentz force. Since pulsed-field magnets are operated close to the destruction limit, the magnet design and choice of materials are of crucial importance.

 \begin{figure}[h]
\begin{center}
\includegraphics[width=0.4\textwidth]{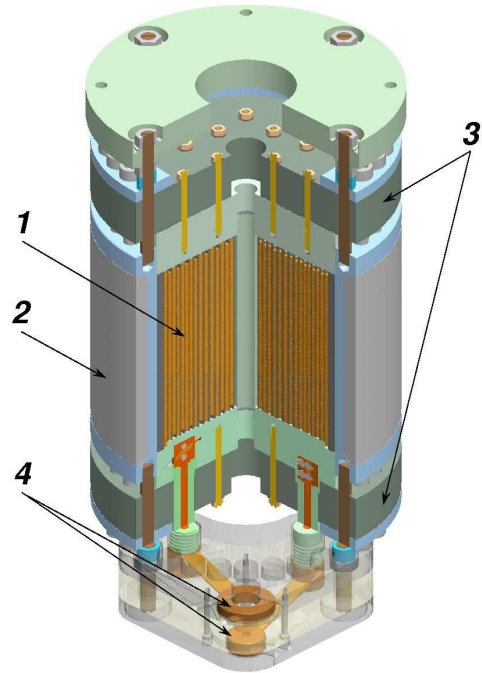}
\caption{\label{fig:Magnet} (Color online) Schematic view of a 8.5 MJ/70 T pulsed-field magnet. Main parts of the magnet are:  coil (1),  reinforcement cylinder (2),  G-10 end flanges (3),  and current leads (4).
}
\end{center}
\end{figure}

Determined by various applications, several types of pulsed-field magnets have been designed and fabricated at the HLD \cite{Coil-1,Coil-2}.  For the ESR experiments  two types of pulsed-field magnets, 8.5 MJ/70 T  and  1.4 MJ/60 T, were used.  The magnet  8.5 MJ/70 T  is used for numerous high-field experiments in fields up to 70 T \cite{Coil-1,Wosnitza}. The coil has a bore with a diameter of 24 mm. The  outer diameter of the coil is 320 mm. A schematic view of the 8.5 MJ/70 T magnet  is shown in Fig.\ \ref{fig:Magnet}. The coil (1) contains 18 conducting layers, each layer has 36 turns. The calculated field homogeneity in the center of the field is better than 7x10$^{-4}$ over 1 cm DSV (diameter spherical volume). High-copper-content alloy wire, Wieland-K88 \cite{Wieland}, with a cross-section of 4x6 mm$^2$ and Kapton-tape insulation is used.  The coil is reinforced by a Zylon-stycast 1266 composite \cite{Zylon} .   The insulated wire is additionally covered  with a braided S2-glass sleeve. A stainless-steel cylinder (2) with  outer diameter of 320 mm is used for external reinforcement of the coil. The G-10 end flanges (3) of the magnet are tightened to the windings with M16 bolts which are screwed into the steel cylinder. A separate assembly provides transitions from copper connectors to the coaxial current leads (4).  The coil inductance is about 15 mH. This magnet has a  pulse-field rise time of 35 ms, while the full pulse duration can be changed  from  approximately 150 to 300 ms  by using different crowbar resistors. A maximal peak current is 23.5 kA \cite{Coil-1}.
As mentioned, during pulses the wire temperature increases from 77 K  to above room temperature.  The typical waiting time between high-field pulses required for temperature relaxation of this coil is about 3 hours. As mentioned, ESR experiments can be performed with smaller, 60 T, coils (1.4 MJ/60 T),  with the bore  diameter of 24 mm and with a  typical full-pulse duration of 40 - 50 ms.  The advantage of using smaller coils is the shorter temperature relaxation time (about 1 hour).

\section{\label{sec:level1}ESR spectrometer}

\begin{figure}
\begin{center}
\includegraphics[width=0.46\textwidth]{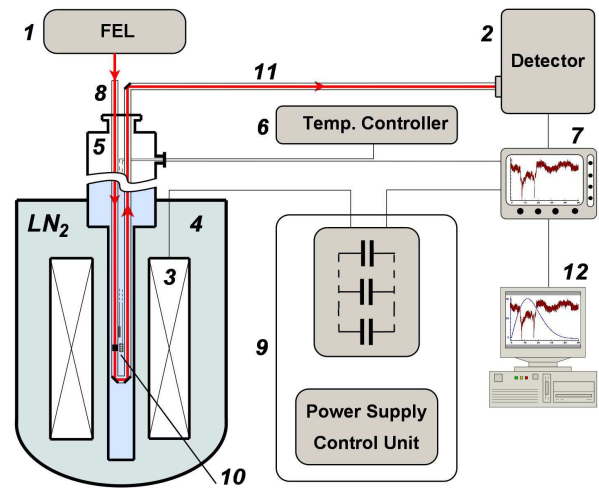}
\vspace{-4 cm}
\caption{\label{fig:Block-scheme} (Color online) Block-diagram of the ESR spectrometer.  The FEL (1) is used  as  source of electromagnetic radiation.   Other parts of the spectrometer are: an IR detector (2), the pulsed-field magnet (3) in a bath with liquid nitrogen (4),  a cryostat (5), a temperature controller (6), a data-acquisition system (7), and a probe (8). The  magnet is fed by the capacitive power supply (9). A temperature sensor and a pick-up coil (10) are installed close to the  sample area. Oversized multi-mode cylindrical waveguides (11) are used to transmit the radiation from the FEL to the probe, and then to the detector.
}
\end{center}
\end{figure}

The block-diagram of the spectrometer is shown in Fig.\ \ref{fig:Block-scheme}. It consists of several parts, including the FEL-based source of electromagnetic radiation (1),  an IR detector (2), the pulsed-field magnet (3), hanging in a bath with liquid nitrogen (4),  a cryostat (5), a temperature controller (6), a data-acquisition system (7), and a probe (8). The pulsed-field magnet is fed by the capacitive power supply (9). A temperature sensor and a pick-up coil (used for measuring the  magnetic field) are installed on the probe (8) close to the sample area (10). The probe is made using oversized multi-mode cylindrical waveguides (11) with inner diameter of 5.6 mm and plane gold-coated mirrors (located at the top and bottom of the probe).  For transmitting the radiation from the probe to the  detector waveguides with inner diameter of 10 mm are used. The spectrometer works in the resonator-free transmission  mode, allowing for ESR experiments  in a broad range of frequencies.

The FEL-based THz source  has been discussed in Section II.  It is important to mention that the frequency range of the FEL spectrometer (1.2 - 75 THz) can be extended by use of conventional   mm- and submm wavelength radiation sources (including BWOs, MVNA, Gunn-diodes, etc.) available at the HLD. Employment of these radiation sources allows for multi-frequency ESR experiments in the frequency range from  30 GHz to 1.3 THz, complementing the ESR facility in the low-frequency range. Superconducting 10 T split-coil and high-homogeneity (10 ppm over 1 cm DSV) 16 T vertical-bore magnets (product of ``Oxford Instruments'') are available for ESR experiments in the  mm- and submm wavelength range as well. In combination with BWOs and MVNA,  the former is used for frequency-domain spectroscopy, while the latter is  employed for tunable-frequency high-resolution  ESR. In addition, an X-band (9 GHz, 1.8 T) Bruker ELEXSYS E500 spectrometer is available for low-frequency ESR experiments where  very high sensitivity and resolution is required.

The broad frequency range of the radiation demands several types of  detectors. These detectors should have  high sensitivity, fast response, and a broad frequency range, suitable for multi-frequency ESR spectroscopy.  For the long  wavelength range of the FEL radiation (down to 200 $\mu$m), a hot-electron magnetically-detuned InSb bolometer (product of ``QMC Instrument Ltd") is used. The detector has a very fast response  (the time constant is of the order of 1 $\mu$s).  For the shorter wavelength range  (down to $\sim$30 $\mu$m) a Ge:Ga photoconducting detector (product of ``QMC Instrument Ltd")  is used.   This detector has a larger time constant, of the order of milliseconds, but is  suitable for ESR experiments in combination with superconducting and mid- and long-pulse magnets.   Both detectors are liquid-helium  cooled and mounted in one cryostat equipped with two optical windows. For operations in the wavelength range of 4 - 30 $\mu$m a liquid-nitrogen cooled HgCdTe (MCT) photovoltaic detector (model  J15D26-M204-S01M-60, product of ``Teledyne Judson Technologies")  with a time constant of 80 ns is used.

A cryostat (``KONTI-IT-Cryostat", product of ``CryoVac") equipped with a variable-temperature inset allows for ESR experiments in a temperature range of 1.4 - 300 K. A ``Lake Shore",  model 340,  temperature controller in combination with standard Cernox$^{TM}$ temperature sensors is used for temperature monitoring and control. A ``Yokogawa", model DL750, digital oscilloscope is used to the register signals from the detector and the pick-up coil.

The magnetic field is measured by integrating the induced signal from the pick-up coil. The ESR signal from DPPH (2,2-diphenyl-1-picrylhydrazyl) was used for calibration of the pick-up  coil installed inside the magnet. A calibration accuracy  better than 0.2$\%$ can be achieved.

 \begin{figure}[h]
\begin{center}
\includegraphics[width=0.5\textwidth]{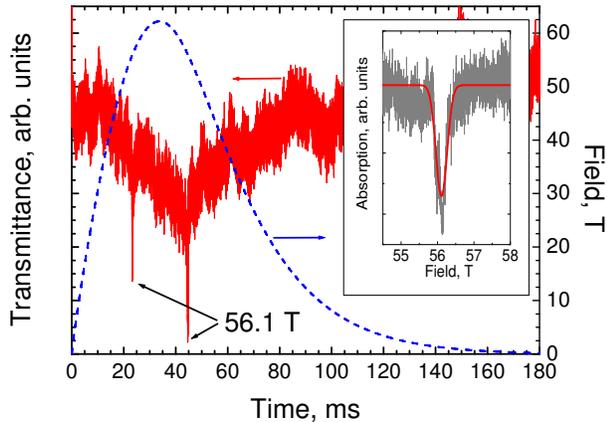}
\caption{\label{fig:DPPH} (Color online) ESR absorption in DPPH (obtained at a wavelength of 190.6 $\mu$m, a temperature of 7 K) and  magnetic field  (shown by the dashed line) vs time.  The maximum of the  field is 62 T. The inset shows the ESR absorption in DPPH as function of  magnetic field. The solid line represents a Gaussian fit with $\Delta B /B\approx0.5\%$. The  8.5 MJ/70 T magnet was used.
}
\end{center}
\end{figure}

The spectral resolution is one of the most important parameters in  ESR spectroscopy, determined by two main factors: (i) homogeneity of the magnetic field and (ii) monochromaticity  of the electromagnetic radiation. Due to the high field homogeneity of the used magnets  (better than 0.1$\%$ over 1 cm DSV) the monochromaticity  of  the electromagnetic radiation should be regarded as the
main parameter determining the spectral resolution of the spectrometer. To demonstrate the high spectral resolution, DPPH was chosen for our  experiments. Due to the  narrow line and the  temperature-independent $g$-factor (2.0036) known with high accuracy, DPPH  is widely used as ESR standard. A powder sample  with a mass of 9 mg was used in this experiment. In  Fig.\ \ref{fig:DPPH}, we show the ESR spectrum obtained for  DPPH at a  wavelength of 190.6 $\mu$m at a temperature of 7 K employing the  8.5 MJ/70 T magnet.  Two very sharp peaks were resolved in the ESR spectrum at 56.1 T during the up and down field sweeps (Fig.\ \ref{fig:DPPH}).  The inset shows the ESR absorption in DPPH as function of  magnetic field. The solid line represents a Gaussian fit. Our analysis reveals a resonance width  at half maximum, $\Delta B=0.3$ T in this measurement,  which corresponds to a spectral resolution $\approx0.5\%$ determined by the   spectral distribution of FEL radiation power due to its bandwidth-limited pulsed nature.

 \begin{figure}[h]
\begin{center}
\includegraphics[width=0.5\textwidth]{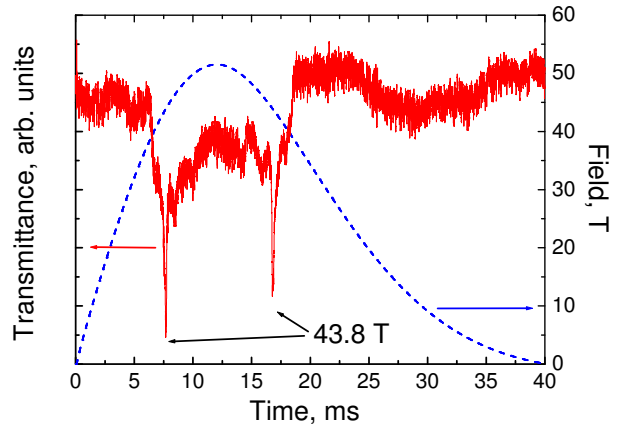}
\caption{\label{fig:6MAP} (Color online)
 ESR absorption in (C$_6$H$_9$N$_2$)CuCl$_3$ (obtained at a wavelength of 228.6 $\mu$m, a temperature of 80 K) and  magnetic field  (shown by the dashed line) vs time. Two sharps absorption lines at 43.8 T corresponding to ESR during the up and down field sweeps are observed. The maximum of the  field is 52 T. The 1.4 MJ/60 T magnet was used.
}
\end{center}
\end{figure}

The ESR spectrum of the  organic quantum spin-1/2 chain compound (C$_6$H$_9$N$_2$)CuCl$_3$ is shown in  Fig.\ \ref{fig:6MAP}. The data were obtained at a wavelength of 228.6 $\mu$m  and at a temperature of 80 K.  A powder sample  with a mass of 16 mg was used in this experiment. Clear and sharp ESR absorption dips corresponding to excitations of exchange-coupled Cu$^{2+}$ centers are visible in the transmitted signal at 43.8 T during the up and down sweeps of the  pulsed field.

Furthermore, the FEL facilities at the FZD have been further successfully  used for studies of the cyclotron resonance in strained  p-InGaAs/GaAs  quantum wells in the  frequency range  of 2.6 - 4.3 THz (69.9 - 115.8 $\mu$m) in magnetic fields up to 55 T \cite{Drachenko} and of the antiferromagnetic resonance in the hexagonal multiferroics YMnO$_3$  \cite{Zvyagin-YMnO}.

\section{\label{sec:level1}Summary}

In conclusion, the feasibility  of using picosecond-pulse  FEL radiation  for the  continuous-wave   ESR spectroscopy has been shown, opening new horizons for high-frequency ESR spectroscopy at high magnetic fields. The new approach is of particular importance for studying magnetic excitation spectra in spin systems with large zero-field splitting (for instance single-molecular magnets, magnetically ordered materials, and  gapped quantum-disordered  antiferromagnets) and  magnetic materials exhibiting field-induced phenomena (such as spin-reorientation transitions, quantum critical behavior, etc.).   The cyclotron resonance will probably remain one of the most important applications of the pulsed-field FEL-spectroscopy in the THz range \cite{Drachenko}. The employment of  FEL for  pulse ESR (sometime called spin-echo ESR) looks  very promising. One of the main difficulties when using picosecond-pulse  FEL radiation  is a lack of fast detectors. The use of electrostatic  FELs with radiation  pulse duration of the order of milliseconds for this kind of spectroscopy seems more attractive \cite{Takahashi}.
Further extension of the magnetic  field range of the spin-echo FEL-based ESR spectroscopy beyond the field limited by superconducting magnets (e.g., using resistive, hybrid and pulsed-field magnets) appears to be very important for studying spin relaxations in a large number of magnetic materials in high magnetic fields.

Apart from studying  resonance phenomena (ESR and cyclotron resonance), the THz-range FEL-based magnetic spectroscopy can have a much broader range of  applications.  In particular, it includes (but is not limited to) frequency-domain dielectric spectroscopy,  ellipsometry, and  polarimetry. These techniques allow for obtaining complex optical conductivity  directly, without using the Kramers-Kronig formalism, and are  of particular importance, for instance, for  studying the superconducting-gap behavior in high-$T_c$ materials. The magnetic field can play here a role of a tuning parameter, closing the superconducting gap and hence suppressing the superconductivity. Complex artificial  optical structures  (metamaterials),  exhibiting unusual optical properties (for instance, negative refraction index) can be the  subject of the FEL spectroscopy as well. FEL pump-probe techniques \cite{pump} in high magnetic fields look very attractive and challenging (particular in pulsed fields) at the same time.  Due to the high radiation power, the FEL light can be used for the investigation of numerous non-linear effects in solids: higher-order harmonic generation, above-threshold ionization, ponderomotive potential effects,  etc. (see for instance \cite{Dekorsy,Sch2,Sch3}). Finally, due to the extraordinary high brightness and tunability,  FEL can be successfully used as an optical source for scattering scanning near-field optical microscopy and spectroscopy  \cite{Schneider,Kehr}.
\\

\section{\label{sec:level1}Acknowledgement}
 We gratefully acknowledge the contribution by F. Pobell, F. Gabriel, and P. Michel  to this project.  We would like to thank M. Helm, O. Drachenko, A. Pronin  for numerous discussions, Y. Skourski for calculations of the coil homogeneity, and the FELBE team for the user support.  The samples of (C$_6$H$_9$N$_2$)CuCl$_3$ were kindly provided by C.P. Landee and M.M. Turnbull. This work was partly supported by the Deutsche Forschungsgemeinschaft, State of Saxony, DeNUF and EuroMagNET II.


\begin{thebibliography}{99}
\bibitem{Zylon} Zylon$^{\circledR}$ is a trademark name of the high-performance fiber poly(p-phenylene-2,6-benzobisoxazole) produced by ``Toyobo Corporation", Japan.
\end{thebibliography}
\end{document}